\documentclass{article}
\usepackage{amsmath,amssymb,graphicx}
\usepackage{textgreek}
\usepackage{authblk}
\usepackage{soul}

\begin{document}

\title{A geometrical view of scalar modulation instability}

\author[1,*]{S. M. Hernandez}
\author[2,3]{P. I. Fierens}
\author[1]{J. Bonetti}
\author[1]{A. D. S\'anchez}
\author[1,3]{D. F. Grosz}
\affil[1]{Grupo de Comunicaciones \'Opticas, Instituto Balseiro,Bariloche, R\'io Negro 8400, Argentina}
\affil[2]{Grupo de Optoelectr\'onica, Instituto Tecnol\'ogico de Buenos Aires,CABA 1106, Argentina}
\affil[3]{Consejo Nacional de Investigaciones Cient\'ificas y T\'ecnicas (CONICET),  Argentina}
\affil[*]{Corresponding author: shernandez@ib.edu.ar}


\date{\today}%
\maketitle

\begin{abstract}

We present a novel approach to the analysis of a full model of scalar modulation instability (MI) by means of a simple geometrical description in the power vs. frequency plane. This formulation allows to relate the shape of the MI gain to any arbitrary dispersion profile of the medium. As a result, we derive a straightforward explanation of the non-trivial dependence of the cutoff power on high-order dispersion and obtain explicitly the power that maximizes the gain. Our approach puts forth a powerful tool to synthesize a desired MI gain with the potential application to a vast number of parametric-amplification and supercontinuum-generation devices whose functioning relies upon modulation instability.

\end{abstract}

\section{Introduction}
\label{sec:intro}

The phenomenon of modulation instability (MI) has been known and
thoroughly studied for many years in a vast number of different areas
of science. In the realm of optical fibers
\cite{Hasegawa.IEEEJournalQuantumElectronics.1980,Anderson.OpticsLett.1984,Tai.PRL.1986,Potasek.OptLett.1987,Potasek.PRA.1987,Nakazawa.PhysRevA.1989,Agrawal.IEEEPhotonicsTechnologyLett.1992},
MI plays a fundamental role as it is intimately connected to the
appearance of optical solitons, which have had a strong impact on
applications to high-capacity fiber-optic communication, among other areas. Modulation
instability also is at the heart of the occurrence of efficient
parametric optical processes heavily relied upon to achieve bright and
coherent light in various spectral ranges. These very same nonlinear
processes are used to provide optical amplification and wavelength
conversion in the telecommunication band, maybe one day enabling
complete photonic control of optical data traffic. In recent years,
nonlinear phenomena such as supercontinuum generation
\cite{Demircan.OpticsComm.2005,Frosz.OptExpress.2006,Frosz.JOSAB.2006,Dudley.OptExpress.2009,Masip.OptLett.2009}
and rogue waves
\cite{Hammani.IEEEPhotonics.2009,Sorensen.JOSAB.2012,Toenger.ScientificReports.2015}
have rekindled the interest in MI.

In the present work we use a complete analysis of MI, first presented in~\cite{Bonetti.PhysRevA.2016} to study coherence in seeded MI, but now turning our attention to the interplay between high-order dispersion and self-steepening.
By analyzing the dependence of the MI gain with pump power in this complete model including both the delayed Raman response and the effect of self-steepening, we find that self-steepening yields an optimum power (in terms of maximizing the gain) and a cutoff power above which the MI gain essentially vanishes, leaving behind only the Raman contribution. These observations regarding the power cutoff and the optimum power in the presence of self-steepening were first reported by Shukla and Rasmussen~\cite{Shukla.OptLett.1986}, and the role of self-steepening was further analyzed by De Angelis \textit{et al.}~\cite{DeAngelis.JOSAB.1996}. However, in both references the effects of high-order dispersion and delayed Raman response were not considered. In this work not only we extend these results, but put forth an analysis in the pump-power-versus-frequency plane, showing that the region where MI gain exists possesses remarkably simple geometrical properties, thus providing a powerful tool for the analysis and synthesis of any arbitrary gain profile.

The remaining of the paper is organized as follows: In Section~\ref{sec:migain}, we briefly review an expression for the MI gain that contemplates all relevant nonlinear effects. Section~\ref{sec:geometry} is devoted to the description of the modulation instability gain in the pump-power-versus-frequency plane and introduces the geometrical model. Analytical expressions for finding gain maxima and the influence of high-order dispersion are presented in Section~\ref{sec:maxima}. Concluding remarks are presented in Section~\ref{sec:conclusions}.

\section{Analytical expression of the MI gain}
\label{sec:migain}

Scalar wave propagation in a lossless nonlinear medium can be described by the
generalized nonlinear Schr\"{o}dinger equation
\cite{Agrawal.NLFO.2012},
\begin{equation}
  \frac{\partial A}{\partial z}-i\hat{\beta}A
  = i \hat{\gamma} A(z,T)\int\limits_{-\infty}^{+\infty}R(T') \left|A(z,T-T')\right|^2 dT',
  \label{eq:gnlse}
\end{equation}
where $A(z,T)$ is the slowly-varying envelope, $z$ is the spatial
coordinate, and $T$ is the time coordinate in a comoving frame at the
group velocity ($=\beta_1^{-1}$). $\hat{\beta}$ and $\hat{\gamma}$ are
operators related to the dispersion and nonlinearity, respectively,
and are defined by
\begin{equation*}
  \hat{\beta}= \sum_{m\geq 2}\frac{i^m}{m!}\beta_m \frac{\partial^m }{\partial T^m},\; \hat{\gamma} = \sum_{n \geq 0} \frac{i^n}{n!} \gamma_n \frac{\partial^n }{\partial T^n}.
\end{equation*}
The $\beta_m$'s are the coefficients of the Taylor expansion of the
propagation constant $\beta(\omega)$ around a central frequency
$\omega_0$. In the convolution integral in the right hand side of
\eqref{eq:gnlse}, $R(T)$ is the response function that
includes both the instantaneous (electronic) and delayed Raman
response of the medium.

The MI gain is given by (see, e.g., \cite{Bonetti.PhysRevA.2016})
\begin{equation}
  \label{eq:migaindef}
  g(\Omega) = 2 \max \{-\mathrm{Im}\{K_1(\Omega)\},-\mathrm{Im}\{K_2(\Omega)\},0\},
\end{equation}
where $\Omega=\omega-\omega_0$, and $K_{1,2}(\Omega)$ are dispersion relations of small perturbations $a = D\exp(iK_{1,2}(\Omega) z)$ to a continuous-wave (CW) pump of frequency $\omega_0$ and power $P_0$ such that $\left(\sqrt{P_0}+a\right) e^{i\gamma_0 P_0 z}$ is an approximate solution to \eqref{eq:gnlse} when only linear terms on the perturbation are considered.


Then, the MI gain with all relevant nonlinear effects present in \eqref{eq:gnlse}  can be obtained (for more details, see Ref.~\cite{Bonetti.PhysRevA.2016}) by finding $K_{1,2}(\Omega)$. In the vast majority of the literature only up to $\gamma_1$ is taken into account. As such, we focus on a simple expression obtained by setting $\gamma_{n \geq 2} = 0$ and $\gamma_1=\gamma_0 \tau_{\mathrm{sh}}$ (accounting for the effect of self-steepening). Then,

\begin{equation}
  \begin{split}
    K_{1,2}(\Omega) = &\tilde{\beta}_o+P_0 \gamma_0 \tau_{\mathrm{sh}}\Omega\left(1+\tilde{R}\right)\pm\\
    \pm&\sqrt{\left(\tilde{\beta}_e+2\gamma_0P_0\tilde{R}\right)\tilde{\beta}_e
      + P_0^2 \gamma_0^2 \tau_{\mathrm{sh}}^2\Omega^2\tilde{R}^2},
  \end{split}
  \label{eq:reldispb}
\end{equation}
with $\tilde{R}$ the Fourier transform of $R$,
\begin{equation*}
  \tilde{\beta}_e(\Omega) = \sum_{n\geq 1} \frac{\beta_{2n}}{(2n)!} \Omega^{2n}, \mbox{ and}\;\tilde{\beta}_o(\Omega) = \sum_{n\geq 1} \frac{\beta_{2n+1}}{(2n+1)!} \Omega^{2n+1}.
\end{equation*}

\section{Geometry of the MI gain}
\label{sec:geometry}

Equations \eqref{eq:migaindef}-\eqref{eq:reldispb} exhibit some
properties of the gain that have been thoroughly studied in the
literature, for instance, the fact that it does not depend on odd
terms of the dispersion relation (e.g., $\beta_3$)
\cite{Potasek.OptLett.1987, Frosz.JOSAB.2006}. However, the derived MI
gain, including the effects of self-steepening and Raman delayed
response, reveals novel aspects related to the self-steepening term
$\gamma_0 \tau_{\mathrm{sh}}$. Indeed, it already has been noted that
this term enables a gain even in a zero-dispersion optical fiber and that, in
general, leads to a narrowing of the MI gain bandwidth
\cite{Abdullaev.OpticsComm.1994,Abdullaev.JOSAB.1997}. These observations are
shown to be a straightforward consequence of the analysis that follows.

It is widely
known (see, e.g., Ref.~\cite{Agrawal.NLFO.2012}) that, for the simplified model that
only takes $\beta_2$ and $\gamma_0$ into account and neglects
self-steepening, as the pump power $P_0$ increases the frequency
$\Omega_{\mathrm{max}}$ where the MI gain attains its maximum, and the
peak gain both increase as, respectively,
\begin{equation}
  \label{eq:maxgainsimple}
  \Omega_{\mathrm{max}} = \pm \sqrt{\frac{2 \gamma_0 P_0}{|\beta_2|}},\qquad g(\Omega_{\mathrm{max}}) = 2 \gamma_0 P_0.
\end{equation}
Enter self-steepening and the relation between the pump power and
the MI gain changes drastically in a non-trivial way, since there
appears an optimum pump power level for which a peak gain is attained, 
and any further increase in pump power makes the MI gain decline. This
relevant observation was first made by Shukla and
Rasmussen~\cite{Shukla.OptLett.1986} with a simplified model of
dispersion expanded up to the GVD parameter. In what follows, we find that this feature is retained when considering an arbitrary number of dispersion terms.  Moreover, we show this to be a corollary of the geometrical properties of the region where MI gain occurs, as defined over the pump-power-versus-frequency plane.

\begin{figure}[t!]
  \includegraphics[width=0.99\linewidth]{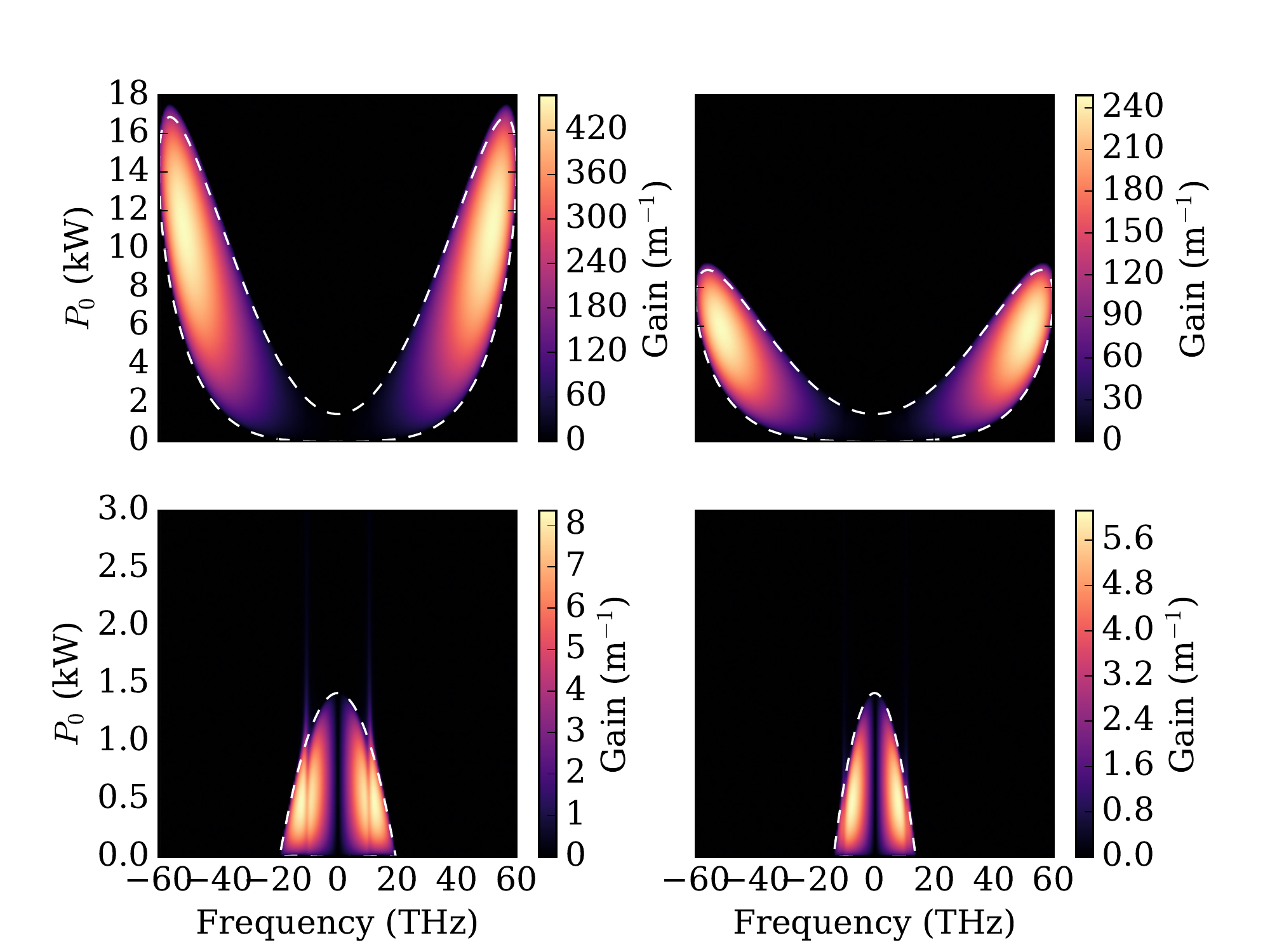}
  \caption{MI gain versus pump power when considering dispersion up to
    $\beta_4$ = -1.6, -0.8, +0.8, +1.6 $\times$ 10\textsuperscript{-3}
    ps\textsuperscript{4}/km (top left, top right, bottom left, and
    bottom right, respectively) including self-steepening. The corresponding MI gain regions are plotted in dashed white.}
  \label{fig:2_2}
\end{figure}

To this purpose, let us analyze the case of Eq.~\eqref{eq:reldispb} where
only the electronic Raman response is taken into account (i.e.,
$\tilde{R}(\Omega)=1$). With the help of some examples (cf. Fig.~\ref{fig:2_2}), this simplification is shown to be not too restrictive. Thus, under this setting,
\begin{equation}
  \label{eq:gain_noraman}
  g(\Omega,P_0) = \left\{
    \begin{array}{cl}
      2 \sqrt{\Delta(\Omega,P_0)} & \mbox{ for } \Delta(\Omega,P_0) > 0 \\
      0 & \mbox{ otherwise,}
    \end{array}
  \right.
\end{equation}
where
\begin{equation}
  \label{eq:delta_WvsP0}
  \Delta(\Omega,P_0) := - P_0^2 \gamma_0^2 \tau_{\mathrm{sh}}^2\Omega^2 
  - P_0 2\gamma_0\tilde{\beta}_e -\tilde{\beta}_e^2.
\end{equation}

Since there is MI gain if, and only if, $\Delta(\Omega,P_0) > 0$,
we may define the \emph{MI gain region} in the $\Omega-P_0$ plane as
\begin{equation}
  \label{eq:MIreg}
  R_{\mathrm{MI}} = \left\{(\Omega,P_0) \in \mathbb{R}\times\mathbb{R}^+:\; \Delta(\Omega,P_0) > 0\right\}. 
\end{equation}
Notice that for $\tau_{\mathrm{sh}} = 0$, we get the usual textbook expression~\cite{Agrawal.NLFO.2012}
\[\begin{split}
\left. R_{\mathrm{MI}}\right|_{\tau_{\mathrm{sh}} = 0} = \left\{\vphantom{\frac{\tilde{\beta}_e(\Omega)}{2\gamma_0}} \right. & (\Omega,P_0) \in
  \mathbb{R}\times \mathbb{R}^+:\; \tilde{\beta}_e(\Omega)<0,\; \\ & \left. P_0 >
  -\frac{\tilde{\beta}_e(\Omega)}{2\gamma_0}\right\},
\end{split}
\]
though we are interested in the case where self-steepening is not
neglected, \textit{i.e.}, $\tau_{\mathrm{sh}} \neq 0$. Here,
$\Delta(\Omega,P_0) = 0$ defines the boundary of $R_{\mathrm{MI}}$ and
is either met when $\Omega = 0$ or whenever $P_0$ is
\begin{equation}
  P_{\pm} = \hat{P}(\Omega)\times \left(1\pm \sqrt{1-\tau_{\mathrm{sh}}^2\Omega^2}\right),
\label{eq:MI_reg_lims}
\end{equation}
where
\begin{equation}
\hat{P}(\Omega) =
-\frac{\tilde{\beta_e}(\Omega)}{\gamma_0\tau_{\mathrm{sh}}^2\Omega^2}.
\label{eq:Pmax}
\end{equation}
Since for each fixed frequency $\partial^2_{P_0} \Delta(\Omega,P_0) <
0$ , it is clear that $\Delta(\Omega,P_0)$ can only be positive
between $P_-$ and $P_+$ when $|\Omega|<\tau_{\mathrm{sh}}^{-1}$.  It is usual to use the approximation $\tau_{\mathrm{sh}}^{-1}\approx \omega_0$, thus neglecting the frequency dependence of the mode effective area \cite{Blow.IEEEQuantumElectronics.1989}. In this case, Eq.~\eqref{eq:MI_reg_lims} limits the frequency to lie in the range $\Omega\in (-\omega_0,\omega_0)$, whereas by taking into account the frequency dependence leads to a slightly increased value of $\tau_{\mathrm{sh}}$ and, hence, to a narrower range of frequencies where the MI gain exists (Ref.~\cite{Kibler.ApplPhysB.2005}). 

\begin{figure}[t]
  \includegraphics[width=0.99\linewidth]{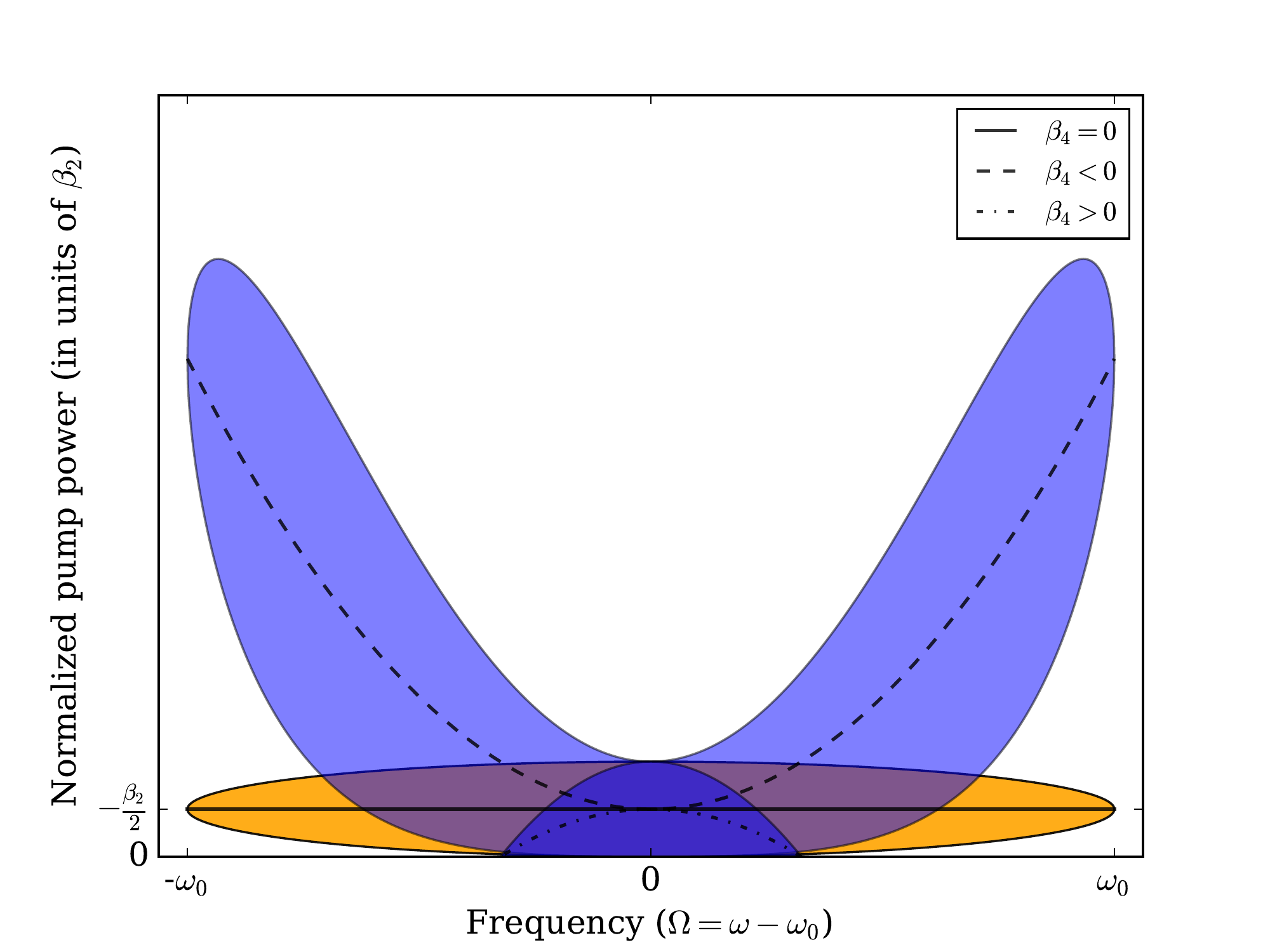}
  \caption{MI gain regions in the plane of normalized pump power
    versus frequency for $\beta_4$ -0.8,0,+0.8 $\times$
    10\textsuperscript{-3} ps\textsuperscript{4}/km. Lines correspond to $\tilde{\beta}_e(\Omega)/\Omega^2$.}
  \label{fig:MI_reg}
\end{figure}

From Eqs.~\eqref{eq:MIreg}-\eqref{eq:Pmax}, we can write
\begin{equation}
  \label{eq:MI_reg_ss}
  \begin{split}
    R_{\mathrm{MI}} = \left\{ \vphantom{\left(\frac{P_0}{\hat{P}(\Omega)} - 1 \right)^2} \right. & (\Omega,P_0) \in [-\omega_0,\omega_0]\times\mathbb{R}^+:\; \tilde{\beta}_e(\Omega) < 0,\; \\ & \left. \left(\frac{P_0}{\hat{P}(\Omega)} - 1 \right)^2 + (\tau_{\mathrm{sh}} \Omega)^2 <  1 \right\}.
  \end{split}
\end{equation}
The $R_{\mathrm{MI}}$ of Eq.~\eqref{eq:MI_reg_ss} has a direct \emph{geometrical interpretation}: since $1 \pm \sqrt{1 - \tau_{\mathrm{sh}}^2\Omega^2}$ defines an ellipse centered at
$(0,1)$ with vertical axis of length 2 and horizontal axis of length $2 \omega_0$, the MI gain region is
given by the portion that lies above the $P_0=0$ axis of the
aforementioned ellipse, bent and stretched along the vertical axis by
$-\tilde{\beta_e}(\Omega)/ (\gamma_0 \tau_{\mathrm{sh}}^2\Omega^2)$. To see this, in Fig.~\ref{fig:MI_reg} we plot MI gain
regions in a plane of normalized power ($P_0 \gamma_0
\tau_{\mathrm{sh}}^2$) versus frequency ($\Omega$) for
$\tilde{\beta_e}(\Omega)= (\beta_2/2) \Omega^2 +(\beta_4/4!)  \Omega^4$
with $\beta_2 = -1\,\mathrm{ps}^2/\mathrm{km}$, $\beta_4$ taking on
the values -0.8, 0, +0.8 $\times$ 10\textsuperscript{-3}
ps\textsuperscript{4}/km, and
$\gamma_0=100\,(\mbox{W-km})^{\mathrm{-1}}$ with a pump centered at a
wavelength of $5$ \textmu m.  Self-steepening is considered by setting
$\tau_{\mathrm{sh}} = \omega_0^{-1}$. The curves
$\tilde{\beta_e}(\Omega)/\Omega^2$ are also plotted as a reference.

With this interpretation in mind we can explain, for instance, the
non-trivial behavior of the power cutoff above which MI gain nearly
vanishes (but for the vestigial contribution due to the
delayed Raman response.) Figure~\ref{fig:2_2} shows the MI gain in
the $\Omega$--$P_0$ plane using the same parameters of
Fig.~\ref{fig:MI_reg} but for $\beta_4$ = -1.6, -0.8, +0.8, +1.6
$\times$ 10\textsuperscript{-3} ps\textsuperscript{4}/km and with the
addition of the delayed Raman response to show its negligible
contribution to the shape of the MI gain region. The Raman response is $R(T) = (1-f_R) \delta(T)+f_R h_R(T)$, with
\begin{equation}
  h_R(T) = \frac{\tau_1^2+\tau_2^2}{\tau_1\tau_2^2}e^{-T/\tau_2}\sin(T/\tau_1) u(T),
  \label{eq:raman}
\end{equation} 
where $u(T)$ is the Heaviside step function, $f_R = 0.031$, $\tau_1 =
15.5\,\mathrm{fs}$, $\tau_2 = 230.5\,\mathrm{fs}$~\cite{Xiong.ApplOpt.2009,Granzow.OptExpress.2012,Karim.OptExpress.2015}.

\begin{figure}[t!]
  \includegraphics[width=0.99\linewidth]{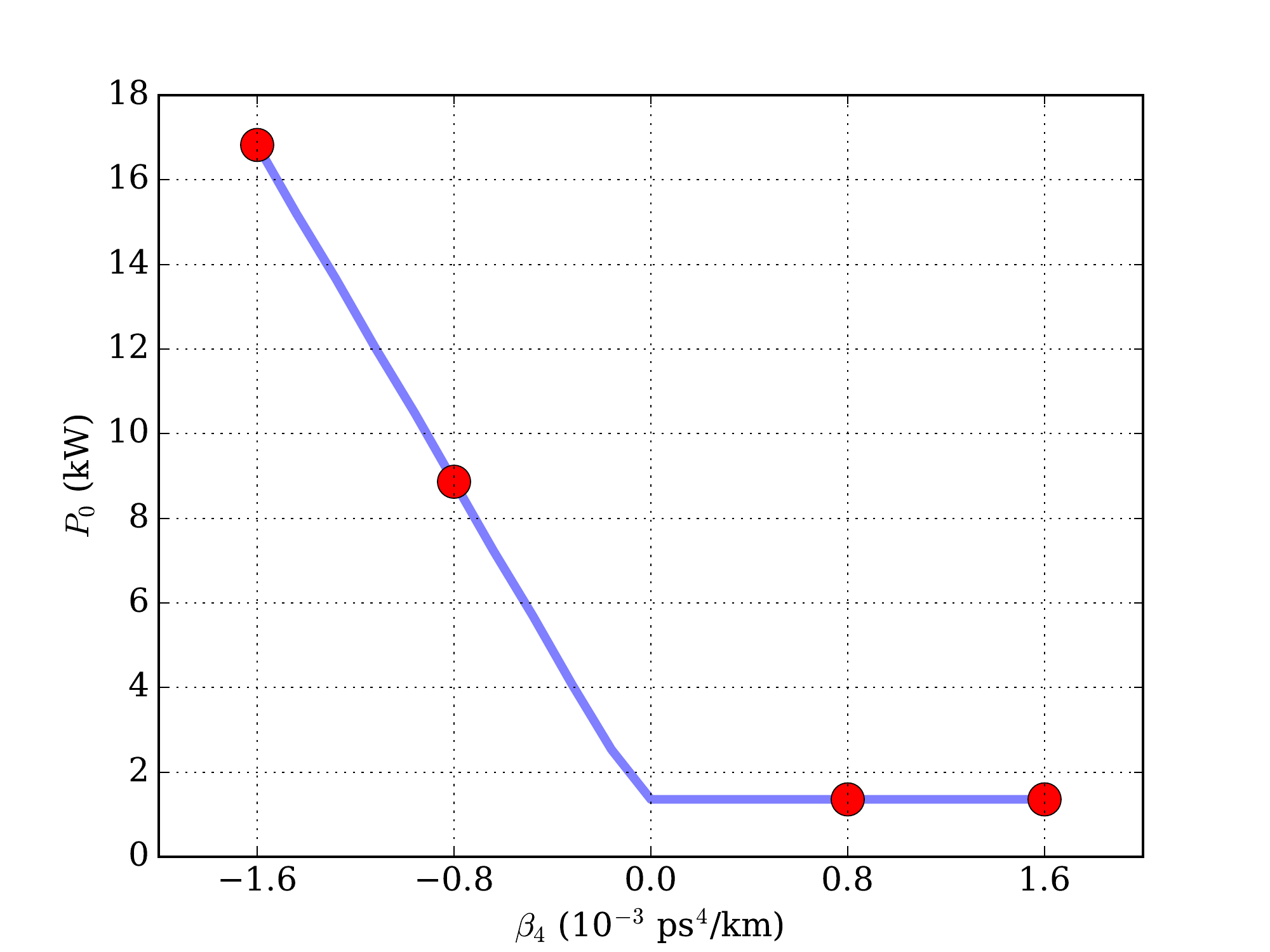}
  \caption{MI gain power cutoff versus $\beta_4$. Dots indicate the
    values of $\beta_4$ used in previous figure.}
  \label{fig:2_3}
\end{figure}

To see what happens with the power cutoff, Fig.~\ref{fig:2_3} shows
the pump power above which MI gain nearly vanishes as a function of
$\beta_4$ for 20 different values ranging from -1.6 to +1.6 $\times$
10\textsuperscript{-3} ps\textsuperscript{4}/km (the $\beta_4$'s of
Fig.~\ref{fig:2_2} are marked in red dots.) As it is readily seen, the
cutoff power varies linearly with $\beta_4$ for $\beta_4 < 0$ and
exhibits a \emph{plateau} when $\beta_4$ contributes towards the
normal dispersion regime (that is, $\beta_4 > 0$).

To explain the \emph{plateau}, note that whenever $\beta_4$ is
positive the ellipse 'bends down', and therefore the power cutoff
remains constant and equal to $-\beta_2/(\gamma_0
\tau_{\mathrm{sh}}^2)$ (\textit{i.e.}, $P_+$ for $\Omega \rightarrow 0$). When
$\beta_4$ is negative, if $|\beta_4|$ is not too small, the cutoff
power (upper limit of the region along the vertical axis) lies near
the position of the maxima of $-\tilde{\beta_e}(\Omega)/\Omega^2$, and
it is easily seen that these maxima vary linearly with $\beta_4$, explaining the approximately linear behavior seen in
Fig.~\ref{fig:2_3} for negative $\beta_4$'s.

All in all, the most obvious and important property that can be
exploited from the geometrical description of the MI region is that,
since the horizontal axis of the ellipse bends with
$-\tilde{\beta_e}(\Omega)/\Omega^2$, one is able to synthesize different MI gain
regions by the arbitrary design of the dispersion profile of the
medium in a straightforward manner.

\section{Location of the MI gain maxima}
\label{sec:maxima}

We may ask for the location of maxima within the MI
gain region as it is paramount to applications which rely upon MI, such as supercontinuum generation from CW lasers and parametric amplification in nonlinear media. In order to do so, we find that
\begin{align}
  \label{eq:derP0}
  \partial_{P_0} \Delta &= -2P_0
  \tau_{\mathrm{sh}}^2\gamma_0^2\Omega^2-2\gamma_0 \tilde{\beta}_e,\\  
  \partial^2_{P_0} \Delta &= -2\tau_{\mathrm{sh}}^2\gamma_0^2\Omega^2.
\end{align}
Since $\partial^2_{P_0}$ is negative definite for $\Omega \neq
0$, by finding zeroes of Eq.~\eqref{eq:derP0} any maximum inside
the MI gain region must have $P_0 = -\tilde{\beta_e}(\Omega)/\gamma_0
\tau_{\mathrm{sh}}^2 \Omega^2 = \hat{P}({\Omega})$. That is, maxima of the modulation instability gain must lie on the lines drawn in Fig.~\ref{fig:MI_reg}.

We may find the location of maxima by differentiating $\Delta(\Omega,P_0)$ with respect to $\Omega$ and proceeding with usual calculus techniques. However, a more intuitive understanding can be reached by defining
\begin{equation}
  \label{eq:gmax_W}
  \begin{split}
  \hat{g}(\Omega) := \max_{P_0} g(\Omega, P_0) & = g(\Omega, \hat{P}(\Omega)) \\ & =  -2 \frac{\tilde{\beta_e}(\Omega)}{\tau_{\mathrm{sh}}|\Omega|}\sqrt{1-\tau_{\mathrm{sh}}^2\Omega^2}
  \end{split}
\end{equation}
for $\tilde{\beta_e}(\Omega) < 0$. It is easy to see that maxima of $g(\Omega,P_0)$ must also be maxima of $\hat{g}(\Omega)$. 

By using Eq.~\eqref{eq:gmax_W} it can be
easily shown that the location of maxima is
$\Omega_{\mathrm{max}}=\pm 1/ 2\tau_{\mathrm{sh}}$ and $P_0 =
-\frac{\beta_2}{2\gamma_0\tau_{\mathrm{sh}}^2}$ for the simple case
where only GVD is considered.  That is, a peak in the MI gain right in the
middle of the power range for which there is gain. This observation
was first reported by Shukla and
Rasmuseen~\cite{Shukla.OptLett.1986}. However, the analysis in
Ref.~\cite{Shukla.OptLett.1986} did not include higher-order
dispersion terms. Thus, if we turn our attention to the influence of
these terms, finding MI gain maxima and their location in the
$\Omega$--$P_0$ plane, which amounts to a simple calculus problem by
means of $\partial_{\Omega} \hat{g}$ and $\partial^2_{\Omega}
\hat{g}$, gives us results which depend parametrically on the dispersion coefficients, and render the analysis (and synthesis) of extrema a straightforward numerical task.

As a simple example, we may consider analyzing the influence of $\beta_4$ in MI gain maxima. If we define $\hat{g}_{\beta_2}(\Omega)$ to be that of Eq.~\eqref{eq:gmax_W} when only GVD is considered, we have that 
\begin{equation*}
\hat{g}(\Omega_{\mathrm{max}}) = \hat{g}_{\beta_2}(\Omega_{\mathrm{max}}) + \partial_{\beta_4} \hat{g}(\Omega_{\mathrm{max}}) \cdot \beta_4
\label{eq:g_b4}
\end{equation*}
given that $\tilde{\beta_e}(\Omega) <0$, $\beta_n = 0$ for $n \geq 6$, where $\pm \Omega_{\mathrm{max}}$ are the arguments that maximize $\hat{g}(\Omega)$. In general, $\Omega_{\mathrm{max}}$ depends on the particular dispersion profile, but it can be shown that, for $|\beta_4|$ large enough, $\Omega_{\mathrm{max}}$ remains nearly constant and $\partial_{\beta_4} \hat{g}(\Omega_{\mathrm{max}})$ also varies little. In practical terms, this means that the gain increase over $\hat{g}_{\beta_2}(\Omega_{\mathrm{max}})$ is proportional to $|\beta_4|$, thus pointing at the strong influence of high-order dispersion (see top left and top right panes of Fig.~\ref{fig:2_2} and note the different scales.)

\section{Conclusions}
\label{sec:conclusions}

In conclusion, we presented a simple geometrical description of a full model of scalar modulation instability. This novel approach allowed us to relate the MI gain profile to any arbitrary dispersion of the medium, and provides a straightforward explanation of the dependence of the cutoff power with high-order dispersion. Further, we showed that the power level maximizing the MI gain is greatly influenced by high-order dispersion and that it can be explicitly obtained.  Finally, the geometrical model can turn into a powerful tool to synthesize a desired MI gain shape with the potential application to a number of parametric-amplification and supercontinuum-generation devices that rely on a precise knowledge of MI dynamics.

\bibliographystyle{unsrt} \bibliography{biblio}

\end{document}